\documentclass{aa}
\usepackage{txfonts}

\usepackage{natbib}
\bibpunct{(}{)}{;}{a}{}{,}

\usepackage{graphicx}

\usepackage{longtable}

\newcommand{\ms}{m\,s$^{-1}$}
\newcommand{\kms}{km\,s$^{-1}$}
\newcommand{\jupmass}{$\rm{M_J}$}
\newcommand{\solmass}{${\rm{M_\odot}}$}
\newcommand{\solrad}{${\rm{R_\odot}}$}
\newcommand{\taugem}{{$\tau$~Gem}}

\begin{document}

\title{Precise radial velocities of giant stars}
\subtitle{V. A brown dwarf and a planet orbiting the K~giant stars \taugem\ and
  91~Aqr\thanks{Based on observations collected at Lick Observatory,
    University of California}\fnmsep\thanks{Based on observations collected at
    the European Southern Observatory, Chile, under program IDs 088.D-0132,
    089.D-0186, and 090.D-0155.}}

\author{D.~S.~Mitchell\inst{1,2}\and 
S.~Reffert\inst{1}\and
T.~Trifonov\inst{1}\and
A.~Quirrenbach\inst{1}\and
D.~A.~Fischer\inst{3}
}

\institute{Landessternwarte, Zentrum f\"{u}r Astronomie der Universität
  Heidelberg, K\"{o}nigstuhl 12, 69117 Heidelberg, Germany
\and Physics Department, California Polytechnic State University, San
  Luis Obispo, CA, 93407, USA\\\email{dsmitche@calpoly.edu}
\and Department of Astronomy, Yale University, New Haven, CT, 06511, USA
}

\date{Received 1 January 2001/Accepted 1 January 2001}

\abstract
{}
{We aim to detect and characterize substellar companions to K~giant stars, to
  further our knowledge of planet formation and stellar evolution of
  intermediate-mass stars.}
{For more than a decade we have used Doppler spectroscopy to acquire high
  precision radial velocity measurements of K~giant stars.  All data for this
  survey have been taken at Lick Observatory.  Our survey includes 373
  G~and K~giants.  Radial velocity data showing periodic variations
  are fitted with Keplerian orbits using a $\chi^2$ minimization technique.}
{We report the presence of two substellar companions to the K~giant stars
  \taugem\ and 91~Aqr.  The brown dwarf orbiting \taugem\ has an orbital
  period of $305.5 \pm 0.1$~days, a minimum mass of 20.6~\jupmass, and an
  eccentricity of $0.031 \pm 0.009$.  The planet orbiting 91~Aqr has an
  orbital period of $181.4 \pm 0.1$~days, a minimum mass of 3.2~\jupmass, and
  an eccentricity of $0.027 \pm 0.026$.  Both companions have exceptionally
  circular orbits for their orbital distance, as compared to all previously
  discovered planetary companions.}
{}


\maketitle

\section{Introduction}

Since the discovery of the first extrasolar planets more than 15 years ago,
there have been over 700 confirmed extrasolar planet
discoveries\footnote{http://www.exoplanets.org}.  Of these planets, more than
60\% have been discovered using the radial velocity (RV) technique, though the
Kepler space telescope has found many more planet candidates that are
waiting confirmation.  Due to their spectral characteristics, solar-type main
sequence stars are the stars most suitable for RV measurements, and so they
have been the targets of the majority of extrasolar planet searches.  However,
a growing number of groups are successfully searching for planets around
evolved giants and subgiants
\citep[e.g.][]{frink02,sato03,seti03,johnjohn,lovis07,niedz07,doel07}.

There are currently 48 known substellar companions orbiting these giant
stars, of which 25 have been published during the past four years.  These
planets allow us to probe more massive stars than are accessible on the main
sequence.  While main sequence planet searches can typically access stars
slightly more massive than our Sun, earlier type stars typically have too few
absorption lines for reliable high-precision RV measurements.  Evolved
stars, such as K~giants, have suitable absorption lines for RV measurements,
but can have much larger masses.  Our sample has typical masses of
1--3~\solmass.  Our results show that red giant stars with masses greater
than 2.7~\solmass\ host very few planets \citep{reffert13}.  K~giant RV
surveys also allow the investigation of how stellar evolution affects
planetary systems.

Our search for planets orbiting giant stars has collected data from 373 stars
for over a decade, with several published planet detections
\citep{frink02,reffert06,aq}.  In this paper we report sub-stellar companions
detected in the RV data of \object{\taugem}\ (HIP~34693) and \object{91~Aqr}
(HIP~114855).  These are both single planet systems at our detection
threshold, and both planets have circular orbits.

In Section~2 we will discuss our observations in detail.  Section~3 contains
information about the stars themselves, including various stellar parameters,
and in Section~4 we derive the planetary orbits.  In Section~5 we present
evidence against the existence of intrinsic stellar causes of the RV
variations.  In Section~6, we discuss the possible multiplicity of the host
stars and the notably low eccentricities of these companions.  Finally, in
Section~7, we summarize our findings.

\section{Observations}

\subsection{Sample Selection}

Our sample of giant stars started with the selection of 86 K~giants brighter
than magnitude~6 in V, showing no signs of duplicity or variability. The
original selection criteria are described in \citet{frink01}.  Observations of
these stars began in June 1999.  In June 2000, 96 additional stars were added
to the sample with less stringent criteria, such as relaxing the criteria
against long-term proper motion and photometric variability.  Three stars were
removed from the sample when it was discovered that they were visual binaries,
leaving a total of 179 stars.  

After the detection of several planets orbiting stars in our original sample
\citep{frink02,reffert06,aq}, 194 G and K~giants were added to the
sample in 2004.  These stars were selected for having higher masses and bluer
colors on average than the original stars.  The larger masses were chosen to
test whether or not more massive stars host more massive planetary
companions.  The color criterion was chosen because bluer stars tend to have
less intrinsic jitter than redder stars.

\subsection{Radial Velocity Data}

The radial velocity observations were taken at Lick Observatory, using the
0.6~m Coud\'{e} Auxiliary Telescope (CAT) with the Hamilton Echelle
Spectrograph \citep{vogt87}.  Our spectra have a resolution of 
$R\,{\approx}\,60\,000$, and cover the wavelength range 3755--9590~{\AA}
(4725--9590~{\AA} before August 2001).  A cell of molecular iodine gas was
heated to $50{\degr}$C and placed in the light path, and the resulting stellar
spectra, with iodine absorption lines, were fitted by models created from
separate iodine and stellar template spectra.  Only the region of the spectrum
with iodine absorption lines (5000--5800~{\AA}) is used for this
procedure. For dwarf stars, this yields Doppler shifts with precisions better
than 3~\ms.  The data acquisition and reduction process is described in more
detail in \citet{butler96}.

We currently have 12--13 years of data for our original set of K~giant stars,
of which both \taugem\ and 91~Aqr are members.  We have 95 RV measurements for
\taugem, and 174 for 91~Aqr spaced over this time period.  The resulting
radial velocities are given in Tables~\ref{tgdata} and~\ref{91aqdata} in
Appendix~B.  Typical exposure times were 15~min for \taugem\ and 12~min for
91~Aqr.  The signal-to-noise ratios for these observations are typically
around 120--150, and the resulting radial velocity measurements have a
median precision of 4.7~\ms\ for \taugem\ and 5.2~\ms\ for 91~Aqr.  Although
this is not the best possible precision available with this method, it is
adequate for the requirements of this project.  Giant stars typically have
larger RV jitter than main sequence stars, on the order of 20~\ms\ or larger
depending on the individual star \citep{hekker06}, so improved precision would
not be beneficial to the detection of a planetary RV signal.


\begin{table}
\caption{Stellar Parameters}
\label{param}
\centering
\begin{tabular}{lcc}
\hline\hline\noalign{\smallskip}
Parameter & \taugem\ & 91~Aqr \\
\hline\noalign{\smallskip}
Spectral type                     & K2III                   & K0III\\
$m_{\rm{v}}$ [mag]\tablefootmark{a} & 4.41                    & 4.24 \\
$M_{\rm{v}}$ [mag]                  & $-0.56 \pm 0.05$       & $0.93 \pm 0.03$ \\
$K$        [mag]\tablefootmark{b} & $1.681 \pm 0.254$       & $1.597 \pm 0.236$\\
$B-V$      [mag]\tablefootmark{a} & $1.261 \pm 0.000$       & $1.107 \pm 0.005$\\
Parallax   [mas]\tablefootmark{a} & $10.16 \pm 0.25$        & $21.78 \pm 0.29$\\
Distance    [pc]                  & $98.4 \pm~^{2.5}_{2.4}$   & $45.9 \pm 0.6$ \\
Radius [\solrad]\tablefootmark{c} & $26.8 \pm 0.7$          & $11.0 \pm 0.1$ \\
$T_{\rm{eff}}$ [K]\tablefootmark{c}  & $4388 \pm 25$           & $4665 \pm 18$ \\
$\log g\ {\rm [cm\,s^{-2}]}$\tablefootmark{c} & $1.96 \pm 0.08$ & $2.52 \pm 0.05$ \\
{}[Fe/H]\tablefootmark{d}         & $0.14 \pm 0.1$          & $-0.03 \pm 0.1$ \\
Mass [\solmass]\tablefootmark{c}  & $2.3 \pm 0.3$           & $1.4 \pm 0.1$ \\
Age        [Gyr]\tablefootmark{c} & $1.22 \pm 0.76$         & $3.56 \pm 0.63$\\
Absolute RV [\kms]\tablefootmark{e} & $22.02 \pm 0.07$ & $-25.75 \pm 0.07$\\
$v\sin i$ [\kms]         & $< 1.0$\tablefootmark{f}& $3.9 \pm 0.5$\tablefootmark{g} \\
\hline
\end{tabular}
\tablefoot{\\
\tablefoottext{a}{Data from Hipparcos:~\citet{hip}}\\
\tablefoottext{b}{Data from 2MASS:~\citet{2mass}}\\
\tablefoottext{c}{\citet{k08}}\\
\tablefoottext{d}{\citet{hekker07}}\\
\tablefoottext{e}{This paper, data acquired using CRIRES, see Section~5.2}\\
\tablefoottext{f}{\citet{dmm99}}\\
\tablefoottext{g}{\citet{toko02}}
}
\end{table}

\section{Stellar Properties}

The stellar properties of both stars are given in Table~\ref{param}.  Visual
magnitude and parallax measurements are taken from Hipparcos observations
\citep{hip}.  K~magnitude values are taken from 2MASS \citep{2mass}.
\citet{hekker07} measured the metallicity of these stars, using our
spectra for their analysis.  \taugem\ is slightly metal-rich, whereas 91~Aqr
has about solar metallicity.

In order to derive stellar parameters such as mass, surface gravity, effective
temperature, stellar radius and stellar age, the evolutionary tracks of
\citet{girardi00} were interpolated onto the observed B-V color, absolute
V magnitude \citep{hip}, and metallicity \citep{hekker07} (trilinear
interpolation). In general, this approach resulted in two possible solutions,
depending on the precise evolutionary status (red giant branch or horizontal
branch) of the star. Probabilities were assigned to each solution, taking the
evolutionary timescale (the speed with which the star moves through that
portion of the evolutionary track) as well as the initial mass functions into
account \citep{k08}.  More details on the method as well as the stellar
parameters for all K giant stars on our Doppler program can be found in
\citet{reffert13}.

Both of these stars were determined to be more likely on the HB by this
method.  \taugem\ was found to have a 64\%\ chance of being on the HB, and
91~Aqr has a 79\%\ chance of being on the HB.  If the stars are RGB stars, the
masses would be $2.5 \pm 0.3$~\solmass\ for \taugem\ and $1.6 \pm
0.2$~\solmass\ for 91~Aqr.

Absolute radial velocity measurements were taken using the CRIRES
spectrograph \citep{crires}.  The $v\sin i$ limit for \taugem\ was measured by
\citet{dmm99}, while the value for 91~Aqr was measured by \citet{toko02}.

\section{Keplerian Orbit}

\begin{figure}
\resizebox{\hsize}{!}{\includegraphics{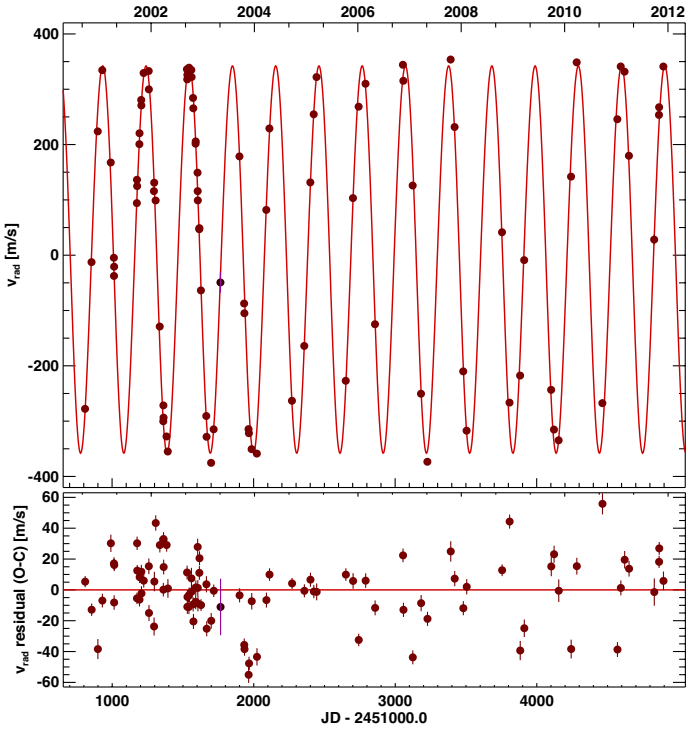}}
\caption{Radial velocity plot for \taugem.  Top panel: RV data with best
  Keplerian fit.  Bottom panel: Residual RV data after subtraction of best fit.}
\label{taugem}
\end{figure}

\begin{figure}
\resizebox{\hsize}{!}{\includegraphics{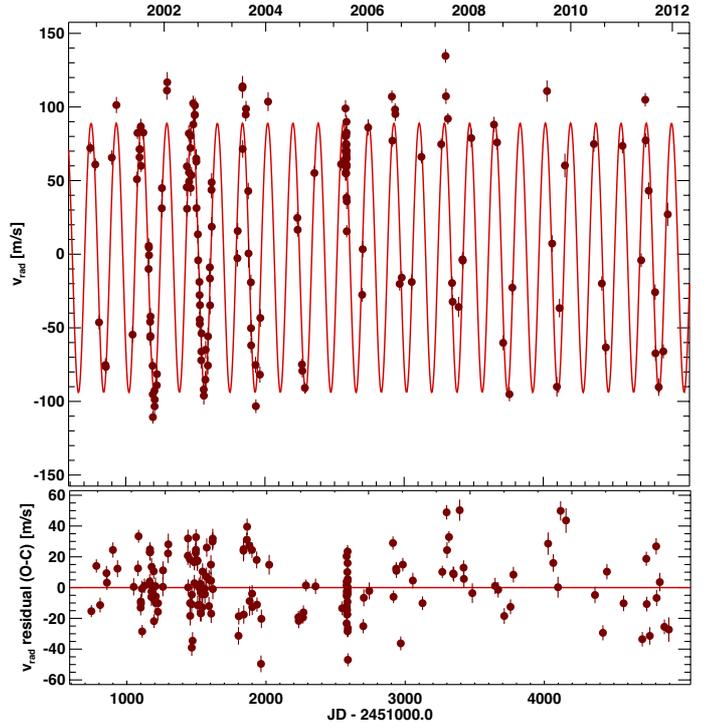}}
\caption{Radial velocity plot for 91~Aqr.  Top panel: RV data with best
  Keplerian fit.  Bottom panel: Residual RV data after subtraction of best fit.}
\label{91aqr}
\end{figure}

The radial velocities of \taugem\ and 91~Aqr are shown in Figures \ref{taugem}
and \ref{91aqr}, along with the best Keplerian fit to the data (upper
panels). The lower panels show the residual radial velocities, after
subtraction of the Keplerian fits. Error bars are included in all plots, but
in many cases are too small to be visible.


For each orbital solution there were six parameters fit to the data: five
orbital parameters and the zero-level of the radial velocities, which was not
constrained by our measurements.  The five orbital parameters are the orbital
period ($P$), the periastron time ($T_0$), the longitude of periastron
($\omega$), the orbital eccentricity ($e$), and the semi-major axis of the
stellar orbit ($a_{1}\sin{i}$).  All of the fitted parameters, with the
exception of the RV zero-point, are included in Table~\ref{orbit}.
Uncertainties shown in those tables are based on the $\chi^2$ fit.  We found
no signature of stellar astrometric motion due to these companions in the
Hipparcos data, and because the threshold for detecting this motion with
Hipparcos is rather high, no meaningful constraints could be placed on either
the companion mass or orbital inclination for either planet.

Fitting orbits with these standard orbital parameters can cause problems when
the orbits have low eccentricities.  For example, the periastron time has very
large errors by definition, if the eccentricity is low.  For this reason we
also fitted the data using the following five orbital parameters: $\log{P}$,
$\log{K_1}$ (where $K_1$ is the semi-amplitude of the RV),
$e\cdot\cos(\omega)$, $e\cdot\sin(\omega)$, and $\lambda_0$ (where $\lambda_0$
is the sum of the longitude of the periastron and the mean anomaly).  Fitting
the data with these parameters resulted in no significant difference to the
orbital solution or the uncertainties.  We also used the revised Lucy-Sweeney
test \citep{lucy} to see whether an upper limit to the eccentricities would be
more appropriate.  While the derived eccentricity of \taugem\ was a sufficient
number of standard deviations away from zero to be accurate, this method
suggests that the eccentricity of 91~Aqr is more appropriately described as
less than 0.034 at the one sigma confidence level, and for the
95\%\ confidence level, less than 0.064.

\begin{table}
\caption{Best Fit Orbital Parameters}
\label{orbit}
\centering
\begin{tabular}{lcc}
\hline\hline\noalign{\smallskip}
Parameter & \taugem\ & 91~Aqr \\
\hline\noalign{\smallskip}

$P$ [days]                   & $305.5 \pm 0.1$        & $181.4 \pm 0.1$       \\
$T_0$ [JD-2450000]           & $3270.7 \pm 14.8$      & $3472.1 \pm 24.5$     \\
$e$                          & $0.031 \pm 0.009$      & $0.027 \pm 0.026$     \\
$\omega$ [\degr]             & $137.4 \pm 0.3$        & $177.3 \pm 0.8$       \\
$a_{1}\sin(i)$ [$10^{-3}$AU]   & $9.83 \pm 0.09$        & $1.53 \pm 0.04$     \\
$f(m)$ [$10^{-8}M_{\sun}]$      & 136                    & 1.44                  \\
$a_2$ [AU]                    & 1.17                   & 0.70                  \\
$K_1$ [\ms]                   & 350.2                  & 91.5                  \\
$m_{2}\sin(i)$ [\jupmass]     & 20.6                   & 3.2                   \\
Reduced $\chi^2$              & 19.8                   & 14.0                  \\
rms scatter around fit [\ms]  & 21.9                   & 18.9                  \\

\hline
\end{tabular}
\end{table}

\begin{figure}
\resizebox{\hsize}{!}{\includegraphics{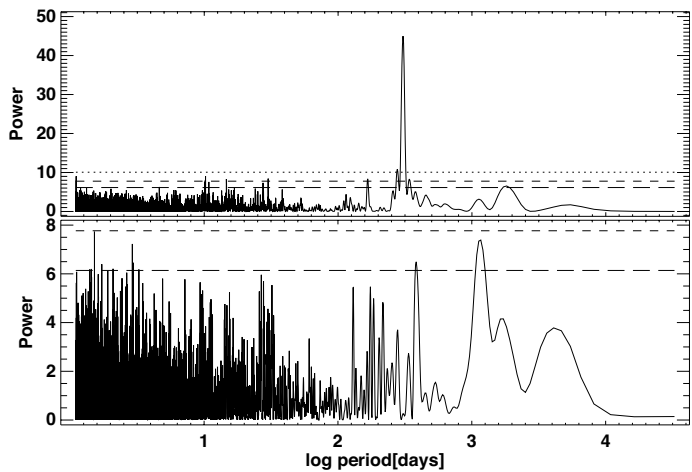}}
\caption{Top: Periodogram of the radial velocity data for \taugem.  There is a
significant peak at 305.5~days, matching the best Keplerian orbital fit.  The
dotted/hyphen/dashed lines show false alarm probabilities of 0.001, 0.01, and
0.05 respectively. Bottom: Periodogram for the residual data after subtraction
of the orbital fit.  There is no significant peak present.}
\label{taugemlosca}
\end{figure}

\begin{figure}
\resizebox{\hsize}{!}{\includegraphics{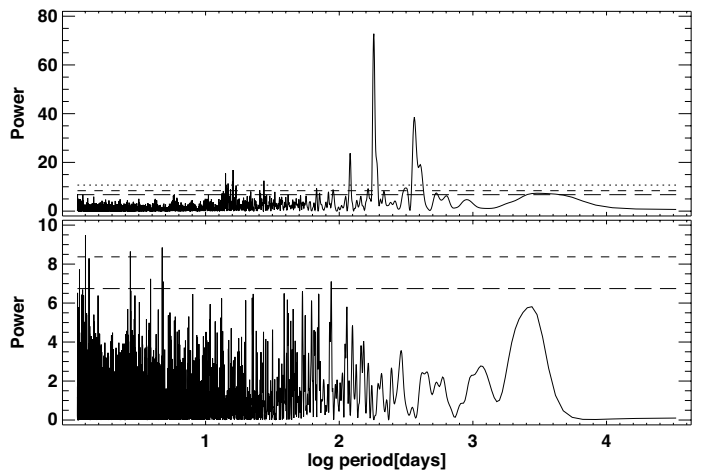}}
\caption{Top: Periodogram of the radial velocity data for 91~Aqr.  There is a
significant peak at 181.4~days, matching the best Keplerian orbital fit.  The
dotted/hyphen/dashed lines show false alarm probabilities of 0.001, 0.01, and
0.05 respectively. Bottom: Periodogram for the residual data after subtraction
of the orbital fit.  There is no significant peak present.}
\label{91aqrlosca}
\end{figure}

Periodograms for our data are shown in Figures~\ref{taugemlosca} and
\ref{91aqrlosca}.  The top panels show the results for the RV data.  Each star
has a single, strong peak at a period matching the Keplerian fit to the data.
91~Aqr has several additional peaks that are weaker but significant.  All of
these peaks are due to aliasing effects.  The two large peaks on either side
of the physical peak are yearly aliases.  To the left of the main peak is a
peak corresponding to the physical frequency $f_0$ added to the annual
frequency $f_{year}$.  To the right is a double peak which combines both $f_0
- f_{year}$ and $f_{year}$ alone.  These two peaks overlap because the orbital
period of this companion is very close to half a year.  At shorter periods
there are smaller peaks at one month and at 14 and 16 days.  These are due to
our observing schedule.  We were usually scheduled for either bi-monthly
observing runs or dark time observing runs every month, thus these aliases
represent our observation frequencies.  All of the aliases seen here are
typical for RV planet searches \citep{dawson10}.

The bottom panels of Figures~\ref{taugemlosca} and \ref{91aqrlosca} show a
periodogram of the data after the Keplerian fit has been removed.  Neither
star has any significant periodicity in the residuals. The residual radial
velocities, after subtraction of the best Keplerian fits, are consistent with
the small intrinsic scatter expected from K-giant stars, at a level of around
20~\ms~\citep{frink01}.

\section{Intrinsic stellar effects}

In addition to planetary or stellar companions, intrinsic stellar
activity can also cause RV variations in giant stars. This makes detections of
substellar companions around K~giants somewhat more challenging than around
inactive stars, since one has to demonstrate that the interpretation of the
observed radial velocity pattern is due to the companion and not due to
intrinsic effects. This has been discussed in detail by \citet{hatzes04}. 

Features on the surface of a star, such as hot or cool spots or plages, can
create asymmetries in the stellar absorption lines \citep{walk92}.  This
asymmetry is seen as a shift in the radial velocity of the star, and as the
star rotates, the Doppler shift will vary periodically.  

taugem\ has a radius of 26.8~\solrad\ and a $v\sin i$ less than
1~\kms.  If one assumes that rotational effects are causing the RV
variations, then the rotation period of the star must be the same as the
period of the radial velocity data, 306~days, leading to a rotational
velocity of 4.4~\kms.  This puts an upper limit on the inclination angle of
the star at 13\fdg0.  With this limit in place, if one assumes starspots
800~K cooler than the stellar surface temperature, the starspots would
need to be larger than 50\%\ of the surface of the star, and there would be
photometric signatures of such a feature many times larger than the
Hipparcos limit of 0.02 magnitudes. Therefore rotational modulation cannot
be the source of RV variability in \taugem.

91~Aqr has a radius of 11.0~\solrad\ and the RV period is 181~days.  This
would imply a rotational velocity of 3.0~\kms, which is less than the
observed rotational velocity of 3.9~\kms.  Given that the uncertainty for
the implied rotational velocity is less than 0.1~\kms and the uncertainty on
the measured rotational velocity is 0.5~\kms, rotational modulation is not a
viable explanation for the 181~day period we observe in 91~Aqr.

Some K giants are known to be pulsating stars. Not all of them pulsate, and
those that do typically display several modes at different frequencies, and with
different amplitudes \citep{hc93}. The pulsations of a star cause parts of the
surface of the star to move toward the observer, and other parts of the surface
to move away from the observer, such that the RV measurements will be affected.
Thus, a particular pulsation frequency in a star may cause RV variations with
the same frequency.

Pulsations can be characterized as radial and non-radial. Radial pulsations
are the simplest pulsation, where the entire star grows and shrinks with a
single frequency. K giant stars typically have radial pulsation periods of a
few days. Using the method of \citet{cks72}, we estimate that the radial
pulsation periods of 91~Aqr should be about 2.6~days, and that of
\taugem\ about 3.0~days. It is clear that the radial pulsation periods for these
stars are several orders of magnitude smaller than the RV periods, ruling
out radial pulsations as the possible cause.

Non-radial pulsations are much more complicated to identify than radial
pulsations.  Any number of periods and amplitudes are possible for the various
modes.  However one would not expect the RV amplitude of the pulsations in the
visible waveband to match the amplitude in the infrared, since the photometric
variations of pulsating stars are different at visible and infrared wavelengths
\citep{percy01}.  On the other hand, if the RV variations are due to a
companion, the IR and visible RV variations should be identical.  Comparing RV
amplitudes at different wavelengths has also proven useful in detecting
intrinsic stellar variations, even in cases where measuring line bisector
variations failed to do so \citep{prato08}.

During 2012 and 2013 we observed both of our stars with the high-resolution IR
spectrograph CRIRES \citep{crires} at the Very Large Telescope (VLT).  CRIRES
spectra have a resolution of $R\,{\approx}\,100\,000$ when used with the 0.2''
slit.  Our spectra cover the wavelength range 1.57--1.61~$\mathrm{\mu m}$,
characterized by many deep and sharp telluric lines used as
references.  To determine RV measurements from these data, we
followed a procedure similar to that of \citet{fig10}.  First we reduced the
data using the standard ESO CRIRES reduction pipeline.  Then we
obtained precise radial velocity measurements by cross-correlating a
synthetic spectrum with the CRIRES spectra.  Our RV measurements using CRIRES
have an estimated uncertainty of 40~\ms. The CRIRES data for both stars are
given in Tables~\ref{tgcrires} and \ref{91acrires} in Appendix~B.  After
collecting the RV measurements from CRIRES, a best fit offset value was
calculated for each of the stars, to fit the IR data to the previously
calculated orbital fit to the visible RV data.

\begin{figure}
\resizebox{\hsize}{!}{\includegraphics{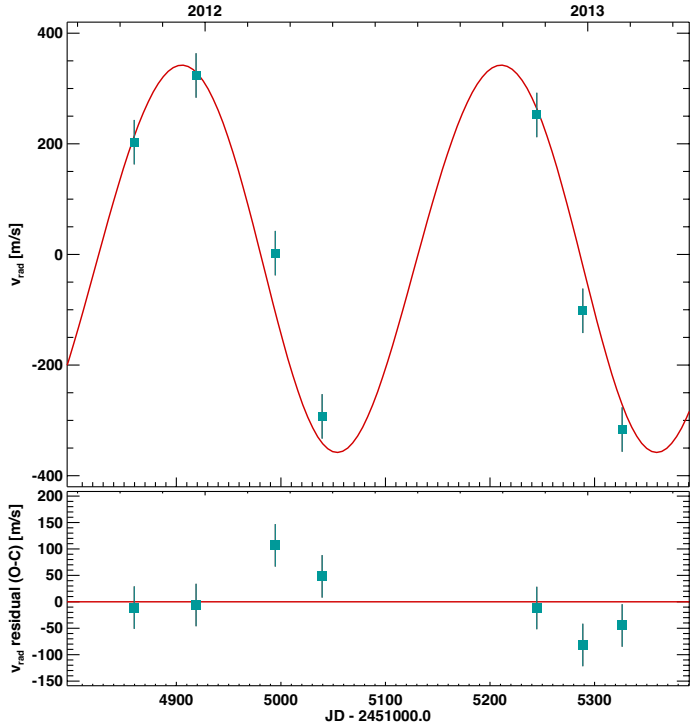}}
\caption{Top panel: \taugem\ CRIRES infrared RV data with orbital fit from
  Lick data. Bottom panel: Residual CRIRES data after subtraction of fit.}
\label{taugemcrires}
\end{figure}

\begin{figure}
\resizebox{\hsize}{!}{\includegraphics{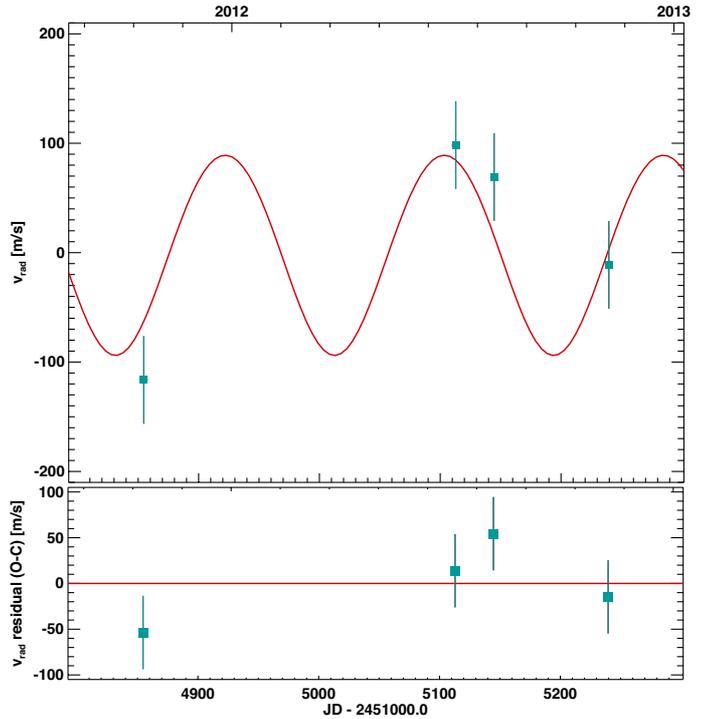}}
\caption{Top panel: 91~Aqr CRIRES infrared RV data with orbital fit from Lick
  data. Bottom panel: Residual CRIRES data after subtraction of fit.}
\label{91aqrcrires}
\end{figure}

Figures~\ref{taugemcrires} and \ref{91aqrcrires} show the CRIRES data.  The top
panels of the figures show the CRIRES data plotted against the Keplerian fit
to the Lick data.  The bottoms of the figures show the residuals of the CRIRES
data after subtraction of the fit.  The scatter around the fit is consistent
with intrinsic jitter of the star.  These plots show that the RV variations in
these stars have the same amplitude and phase in both visible and IR
bandpasses, confirming that the source of the variations are the companions
and not intrinsic variations.

\section{Discussion}

\subsection{Stellar multiplicity}

Both of the stars discussed in this paper are listed in the literature as
being multiple stars.  \taugem\ is indicated in the Washington Double Star
catalog \citep[WDS;][]{wds} as having two companion stars.  The primary
companion was measured to be separated by 2\arcsec, and has a visual magnitude
of $m_{\rm{v}}=11.0$ mag.  More recent observation by Hipparcos did not detect
this secondary component. This component is most likely physically connected
to our star, based on the very small separation.  At the Hipparcos distance of
98.4~pc, the secondary star would be a K0~dwarf, separated by about 187~AU.

For a star with a companion of known mass and separation, one can
find the maximum rate of change for the radial velocity of any orbit to be

\begin{equation}
a_{z_{max}}=\frac{2Gm_{c}\varpi^{2}}{3\sqrt{3}\rho^{2}}
\end{equation}

where $G$ is the gravitational constant, $m_{c}$ is the companion mass,
$\varpi$ is the parallax for the stars and $\rho$ is the angular
separation between   the stars.  The derivation of this equation is given in
Appendix~A. If we   assume that the companion here is a K0~dwarf of mass
0.8~\solmass, then the   maximum radial velocity change would be
1.6~$\rm{m/s/yr}$, which could be   detectable with our current data.
However, many orbits with RV variations   below our detection threshold are
also possible.

An additional companion is listed in the
WDS, at a separation of about 59\arcsec, and a magnitude of $m_{\rm{v}}=12.4$
mag, though the physical connection of this component is dubious. This would
correspond to a physical separation of well over 5000~AU, which would not be
visible in our RV data.

91~Aqr is listed in the WDS as a member of a quintuple star system.  The
primary companion to 91~Aqr is a visual binary pair of magnitudes
10.5 and 10.7, separated from the primary component by 50\arcsec. At the
Hipparcos distance of 45.9~pc, this would lead to a separation of over
2000~AU, too large to be visible in the RV data. \citet{rag06} show that the
primary companion to 91~Aqr is a binary star pair by matching the proper
motion and radial velocities of the two components, and by comparing
spectroscopic parallaxes.

\citet{zirm07} derived the orbit of the BC components as having an
orbital period of 83.6~yr, a semi-major axis of $0\farcs47$, an eccentricity
of 0.45, and an inclination of 87\degr, which is nearly edge-on.  This
semi-major axis corresponds to 21.5~AU.

\citet{rag06} also show that the other two companions listed in the WDS are in
fact field stars. Thus, 91~Aqr~b is one of a handful of planets in a triple
star system.

\subsection{Brown dwarf companions}

It is well established that Sun-like main sequence stars are much more likely
to have planetary mass companions than brown dwarf companions
\citep{marcy00}.  From the California \&\ Carnegie Planet Search and the
McDonald Observatory Planet Search, the frequency at which we find brown dwarf
companions orbiting Sun-like main sequence stars with orbital periods less
than five years is less than 1\%\ \citep[e.g.][]{vogt02,patel07,witt09}. By
controlling for selection effects, \citet{grether06} approximate the fraction
of Sun-like stars harboring giant planet companions with orbits less than 5
years to be $5 \pm 2\%$, and the fraction of such stars harboring brown dwarf
companions at close orbits to be less than 1\%. Since they also calculated
that $11 \pm 3\%$ of Sun-like stars have stellar companions at close orbits,
this demonstrates the relative paucity of companions in the brown dwarf mass
regime, compared to both larger and smaller masses.

It appears that giant stars may host more brown dwarfs than solar-type main
sequence stars.  To date, 10 companions in the brown dwarf mass
regime have been found around giant stars, as compared to nine brown dwarfs
orbiting main sequence stars and two orbiting subgiants.  Three of the nine
main sequence companions were discovered with transit observations.  This is
all true despite the giant stars being more massive, on average, than the
stars in the main sequence samples, making RV detections more difficult due to
the smaller amplitudes, and despite the giant star surveys having a smaller
stellar sample.  It is difficult to estimate the rate of brown dwarf detection
in giant stars, due to the less complete nature of these surveys, and the large
and varying amount of stellar jitter in these stars.  However, it seems likely
that brown dwarfs are not quite so rare around the more massive, evolved stars
of the giant surveys, as these more massive stars would have had more massive
protoplanetary disks from which to form companions.

\subsection{Eccentricity}

\begin{figure}
\resizebox{\hsize}{!}{\includegraphics{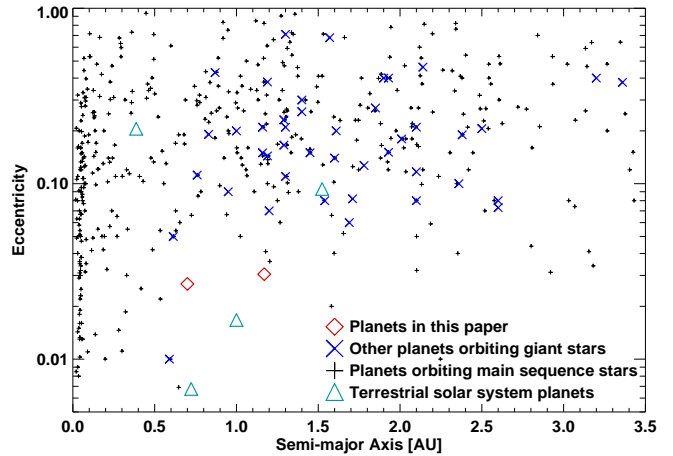}}
\caption{Eccentricities for substellar companions.  Companions
  from this paper are shown as diamonds, all other companions of giant stars
  are shown with a cross.  Companions to main sequence stars are shown as small
  plusses.  Terrestrial planets from our solar system are shown as triangles.}
\label{ecc}
\end{figure}

Both of the planetary orbits described here are extremely circular.  In fact,
they could be considered the most circular orbits found among planets orbiting
at this distance from their host star.  Figure~\ref{ecc} shows the orbital
eccentricity of known extrasolar planets plotted against the semi-major
axis of their orbits.  Symbols for the different categories of planets are
given in the plot.  Planets with orbits greater than 3.5~AU are omitted, as
are planets without properly measured values and uncertainties for the orbital
eccentricity.  Orbital values for the companions to Pollux and $\nu$~Oph are
taken from our data, since it is the most extensive RV data set for these
stars, with the best available orbital fits. There are, however, other
published values \citep{hatzes06,sato12}. In particular, the alternate
published value for the companion to Pollux lists an eccentricity of $0.02 \pm
0.03$, as opposed to our current best fit value of $0.05 \pm 0.01$.

From the plot it is clear that one must look at closer orbits to find planets
with as circular orbits as the planets described here, with a few possible
exceptions.  One other K~giant companion has a lower published eccentricity,
\citep[HD~32518~b,][]{doel09}, but the value for that eccentricity is
questionable.  The published value for HD~32518~b is $e = 0.01 \pm 0.03$, which
according to the Lucy-Sweeney test should be replaced by an upper limit on the
eccentricity of 0.06 for the 95\%\ confidence level, and at the one sigma
level the upper limit would be 0.03.  This is approximately the same value as
the two orbits presented in this paper, but with larger uncertainty.

A more interesting planet with a more circular orbit than those presented
here has an eccentricity of $0.0069 \pm~^{0.0010}_{0.0015}$ and semi-major
axis of 0.65~AU \citep{doyle11}.  This is the famous circumbinary planet
Kepler~16~b.  A second interesting planet with a similar orbit
(though larger semi-major axis) is HD~60532~c \citep{desort08}.  This is the
outer of two planets orbiting this star, orbiting at 1.58~AU with an
eccentricity of $0.02 \pm 0.02$.  Again, using the Lucy-Sweeney test, this
might more accurately be considered an orbit with an upper limit to the
eccentricity of 0.05 for the 95\%\ confidence level, and 0.03 for the one
sigma confidence level.  It is also interesting to note that the authors of
this discovery believe there to be significant interaction between these two
planets, including a possible 3-to-1 orbital resonance, with variations in the
semi-major axis and eccentricity of the orbits.

Finally, one last notable planet is one with a semi-major axis of 2.25~AU, and
an eccentricity of $0.01 \pm 0.03$: HD~159868~c \citep{witt12}.  Like
HD~60532~c, this is the outer planet of a two-planet system where interactions
are possible, though the orbits are stable.  Like HD~32518~b, the uncertainty
on the eccentricity is so high that an upper limit of 0.03 (for one sigma
confidence) is appropriate.


Several groups have investigated the theoretical effects of stellar evolution
on the orbits of planets \citep{vill09,kuni11}.  Though these studies
have focused on the change in semi-major axis of the planet, and the
possibility of engulfment by the RGB star, presumably planets just outside the
engulfment limit will still be strongly affected by the star's tidal
forces. Effects of this interaction might include orbital circularization.
Another possibility is that planets beyond a critical distance may evolve into
orbits with a larger semi-major axis due to mass loss by the host
star.  Even though one is far more likely to find circular orbits for
closely orbiting planets, either of these phenomena could explain more distant
orbits, such as those described here, having very low eccentricities.  It is
interesting that the only planets discovered with similar orbits to the
planets discussed in this paper are in multiple planetary systems that likely
interact, orbit a binary star, or are orbiting evolved stars.

\section{Summary}

In this paper we report the detection of two substellar companions orbiting
the K~giant stars \taugem\ and 91~Aqr, using the Hamilton spectrograph at Lick
Observatory.  The companion to \taugem\ has a minimum mass of
20.6~\jupmass, putting it in the brown dwarf regime.  The companion
to 91~Aqr has a minimum mass of 3.2~\jupmass, putting it in the giant
planet mass range.  Both companions have extremely circular orbits, especially
considering their relatively distant orbits, and the orbital characteristics
have been stable over more than 12 years of observation.  RV data taken in the
IR match the orbital fit, confirming the presence of a companion as the source
of the RV variations.

\begin{acknowledgements}

We would like to thank the staff at Lick Observatory for their support over
the years of this project.  We would like to thank the CAT observers that
assisted with this project, including Saskia Hekker, Simon Albrecht, David
Bauer, Christoph Bergmann, Stanley Browne, Kelsey Clubb, Dennis K{\"u}gler,
Christian Schwab, Julian St{\"u}rmer, Kirsten Vincke, and Dominika
Wylezalek. We graciously acknowledge the work of Geoffrey Marcy in creating
the iodine cell system at Lick Observatory, and writing the original Doppler
code. We thank Mathias Zechmeister and Ansgar Reiners for their help with the
acquisition and reduction of the CRIRES data.  This research has made use of
the SIMBAD database and the VizieR catalog access tool, CDS, Strasbourg,
France.  This publication makes use of data products from the Two Micron All
Sky Survey, which is a joint project of the University of Massachusetts and
the Infrared Processing and Analysis Center/California Institute of
Technology, funded by the National Aeronautics and Space Administration and
the National Science Foundation.  This research has made use of the Washington
Double Star Catalog maintained at the U.S. Naval Observatory.  This research
has made use of the Exoplanet Orbit Database and the Exoplanet Data Explorer
at exoplanets.org.

\end{acknowledgements}

\bibliographystyle{aa}
\bibliography{planets}

\begin{thebibliography}{47}
\expandafter\ifx\csname natexlab\endcsname\relax\def\natexlab#1{#1}\fi

\bibitem[{{Butler} {et~al.}(1996){Butler}, {Marcy}, {Williams}, {McCarthy},
  {Dosanjh}, \& {Vogt}}]{butler96}
{Butler}, R.~P., {Marcy}, G.~W., {Williams}, E., {et~al.} 1996, \pasp, 108, 500

\bibitem[{{Cox} {et~al.}(1972){Cox}, {King}, \& {Stellingwerf}}]{cks72}
{Cox}, J.~P., {King}, D.~S., \& {Stellingwerf}, R.~F. 1972, \apj, 171, 93

\bibitem[{{Dawson} \& {Fabrycky}(2010)}]{dawson10}
{Dawson}, R.~I. \& {Fabrycky}, D.~C. 2010, \apj, 722, 937

\bibitem[{{de Medeiros} \& {Mayor}(1999)}]{dmm99}
{de Medeiros}, J.~R. \& {Mayor}, M. 1999, \aaps, 139, 433

\bibitem[{{Desort} {et~al.}(2008){Desort}, {Lagrange}, {Galland}, {Beust},
  {Udry}, {Mayor}, \& {Lo Curto}}]{desort08}
{Desort}, M., {Lagrange}, A.-M., {Galland}, F., {et~al.} 2008, \aap, 491, 883

\bibitem[{{D{\"o}llinger} {et~al.}(2009){D{\"o}llinger}, {Hatzes}, {Pasquini},
  {Guenther}, \& {Hartmann}}]{doel09}
{D{\"o}llinger}, M.~P., {Hatzes}, A.~P., {Pasquini}, L., {Guenther}, E.~W., \&
  {Hartmann}, M. 2009, \aap, 505, 1311

\bibitem[{{D{\"o}llinger} {et~al.}(2007){D{\"o}llinger}, {Hatzes}, {Pasquini},
  {Guenther}, {Hartmann}, {Girardi}, \& {Esposito}}]{doel07}
{D{\"o}llinger}, M.~P., {Hatzes}, A.~P., {Pasquini}, L., {et~al.} 2007, \aap,
  472, 649

\bibitem[{{Doyle} {et~al.}(2011){Doyle}, {Carter}, {Fabrycky}, {Slawson},
  {Howell}, {Winn}, {Orosz}, {Pr\v{s}a}, {Welsh}, {Quinn}, {Latham}, {Torres},
  {Buchhave}, {Marcy}, {Fortney}, {Shporer}, {Ford}, {Lissauer}, {Ragozzine},
  {Rucker}, {Batalha}, {Jenkins}, {Borucki}, {Koch}, {Middour}, {Hall},
  {McCauliff}, {Fanelli}, {Quintana}, {Holman}, {Caldwell}, {Still},
  {Stefanik}, {Brown}, {Esquerdo}, {Tang}, {Furesz}, {Geary}, {Berlind},
  {Calkins}, {Short}, {Steffen}, {Sasselov}, {Dunham}, {Cochran}, {Boss},
  {Haas}, {Buzasi}, \& {Fischer}}]{doyle11}
{Doyle}, L.~R., {Carter}, J.~A., {Fabrycky}, D.~C., {et~al.} 2011, Science,
  333, 1602

\bibitem[{{Figueira} {et~al.}(2010){Figueira}, {Pepe}, {Melo}, {Santos},
  {Lovis}, {Mayor}, {Queloz}, {Smette}, \& {Udry}}]{fig10}
{Figueira}, P., {Pepe}, F., {Melo}, C.~H.~F., {et~al.} 2010, \aap, 511, A55

\bibitem[{{Frink} {et~al.}(2002){Frink}, {Mitchell}, {Quirrenbach}, {Fischer},
  {Marcy}, \& {Butler}}]{frink02}
{Frink}, S., {Mitchell}, D.~S., {Quirrenbach}, A., {et~al.} 2002, \apj, 576,
  478

\bibitem[{{Frink} {et~al.}(2001){Frink}, {Quirrenbach}, {Fischer}, {R{\"o}ser},
  \& {Schilbach}}]{frink01}
{Frink}, S., {Quirrenbach}, A., {Fischer}, D., {R{\"o}ser}, S., \& {Schilbach},
  E. 2001, \pasp, 113, 173

\bibitem[{{Girardi} {et~al.}(2000){Girardi}, {Bressan}, {Bertelli}, \&
  {Chiosi}}]{girardi00}
{Girardi}, L., {Bressan}, A., {Bertelli}, G., \& {Chiosi}, C. 2000, \aaps, 141,
  371

\bibitem[{{Grether} \& {Lineweaver}(2006)}]{grether06}
{Grether}, D. \& {Lineweaver}, C.~H. 2006, \apj, 640, 1051

\bibitem[{{Hatzes} \& {Cochran}(1993)}]{hc93}
{Hatzes}, A.~P. \& {Cochran}, W.~D. 1993, \apj, 413, 339

\bibitem[{{Hatzes} {et~al.}(2006){Hatzes}, {Cochran}, {Endl}, {Guenther},
  {Saar}, {Walker}, {Yang}, {Hartmann}, {Esposito}, {Paulson}, \&
  {D{\"o}llinger}}]{hatzes06}
{Hatzes}, A.~P., {Cochran}, W.~D., {Endl}, M., {et~al.} 2006, \aap, 457, 335

\bibitem[{{Hatzes} {et~al.}(2004){Hatzes}, {Setiawan}, {Pasquini}, \& {da
  Silva}}]{hatzes04}
{Hatzes}, A.~P., {Setiawan}, J., {Pasquini}, L., \& {da Silva}, L. 2004, in ESA
  Special Publication, Vol. 538, Stellar Structure and Habitable Planet
  Finding, ed. F.~{Favata}, S.~{Aigrain}, \& A.~{Wilson}, 87--92

\bibitem[{{Hekker} \& {Mel{\'e}ndez}(2007)}]{hekker07}
{Hekker}, S. \& {Mel{\'e}ndez}, J. 2007, \aap, 475, 1003

\bibitem[{{Hekker} {et~al.}(2006){Hekker}, {Reffert}, {Quirrenbach},
  {Mitchell}, {Fischer}, {Marcy}, \& {Butler}}]{hekker06}
{Hekker}, S., {Reffert}, S., {Quirrenbach}, A., {et~al.} 2006, \aap, 454, 943

\bibitem[{{Johnson} {et~al.}(2006){Johnson}, {Marcy}, {Fischer}, {Henry},
  {Wright}, {Isaacson}, \& {McCarthy}}]{johnjohn}
{Johnson}, J.~A., {Marcy}, G.~W., {Fischer}, D.~A., {et~al.} 2006, \apj, 652,
  1724

\bibitem[{{K\"{a}ufl} {et~al.}(2004){K\"{a}ufl}, {Ballester}, {Biereichel},
  {Delabre}, {Donaldson}, {Dorn}, {Fedrigo}, {Finger}, {Fischer}, {Franza},
  {Gojak}, {Huster}, {Jung}, {Lizon}, {Mehrgan}, {Meyer}, {Moorwood}, {Pirard},
  {Paufique}, {Pozna}, {Siebenmorgen}, {Silber}, {Stegmeier}, \&
  {Wegerer}}]{crires}
{K\"{a}ufl}, H.-U., {Ballester}, P., {Biereichel}, P., {et~al.} 2004, in
  Society of Photo-Optical Instrumentation Engineers (SPIE) Conference Series,
  Vol. 5492, Society of Photo-Optical Instrumentation Engineers (SPIE)
  Conference Series, ed. A.~F.~M. {Moorwood} \& M.~{Iye}, 1218--1227

\bibitem[{{Kunitomo} {et~al.}(2011){Kunitomo}, {Ikoma}, {Sato}, {Katsuta}, \&
  {Ida}}]{kuni11}
{Kunitomo}, M., {Ikoma}, M., {Sato}, B., {Katsuta}, Y., \& {Ida}, S. 2011,
  \apj, 737, 66

\bibitem[{{K\"unstler}(2008)}]{k08}
{K\"unstler}, A. 2008, Massen-und Altersbestimmung einer Auswahl von G- und
  K-Riesensternen, Diplomarbeit, Landessternwarte, Heidelberg University,
  Germany

\bibitem[{{Lovis} \& {Mayor}(2007)}]{lovis07}
{Lovis}, C. \& {Mayor}, M. 2007, \aap, 472, 657

\bibitem[{{Lucy}(2013)}]{lucy}
{Lucy}, L.~B. 2013, \aap, 551, A47

\bibitem[{{Marcy} \& {Butler}(2000)}]{marcy00}
{Marcy}, G.~W. \& {Butler}, R.~P. 2000, \pasp, 112, 137

\bibitem[{{Mason} {et~al.}(2001){Mason}, {Wycoff}, {Hartkopf}, {Douglass}, \&
  {Worley}}]{wds}
{Mason}, B.~D., {Wycoff}, G.~L., {Hartkopf}, W.~I., {Douglass}, G.~G., \&
  {Worley}, C.~E. 2001, \aj, 122, 3466

\bibitem[{{Niedzielski} {et~al.}(2007){Niedzielski}, {Konacki}, {Wolszczan},
  {Nowak}, {Maciejewski}, {Gelino}, {Shao}, {Shetrone}, \& {Ramsey}}]{niedz07}
{Niedzielski}, A., {Konacki}, M., {Wolszczan}, A., {et~al.} 2007, \apj, 669,
  1354

\bibitem[{{Patel} {et~al.}(2007){Patel}, {Vogt}, {Marcy}, {Johnson}, {Fischer},
  {Wright}, \& {Butler}}]{patel07}
{Patel}, S.~G., {Vogt}, S.~S., {Marcy}, G.~W., {et~al.} 2007, \apj, 665, 744

\bibitem[{{Percy} {et~al.}(2001){Percy}, {Wilson}, \& {Henry}}]{percy01}
{Percy}, J.~R., {Wilson}, J.~B., \& {Henry}, G.~W. 2001, \pasp, 113, 983

\bibitem[{{Prato} {et~al.}(2008){Prato}, {Huerta}, {Johns-Krull}, {Mahmud},
  {Jaffe}, \& {Hartigan}}]{prato08}
{Prato}, L., {Huerta}, M., {Johns-Krull}, C.~M., {et~al.} 2008, \apjl, 687,
  L103

\bibitem[{{Quirrenbach} {et~al.}(2011){Quirrenbach}, {Reffert}, \&
  {Bergmann}}]{aq}
{Quirrenbach}, A., {Reffert}, S., \& {Bergmann}, C. 2011, in American Institute
  of Physics Conference Series, Vol. 1331, American Institute of Physics
  Conference Series, ed. S.~{Schuh}, H.~{Drechsel}, \& U.~{Heber}, 102--109

\bibitem[{{Raghavan} {et~al.}(2006){Raghavan}, {Henry}, {Mason}, {Subasavage},
  {Jao}, {Beaulieu}, \& {Hambly}}]{rag06}
{Raghavan}, D., {Henry}, T.~J., {Mason}, B.~D., {et~al.} 2006, \apj, 646, 523

\bibitem[{{Reffert} {et~al.}(in prep.){Reffert}, {Bergmann}, {Quirrenbach},
  {Trifonov}, {K\"{u}nstler}, {Fischer}, \& {Marcy}}]{reffert13}
{Reffert}, S., {Bergmann}, C., {Quirrenbach}, A., {et~al.} in prep.

\bibitem[{{Reffert} {et~al.}(2006){Reffert}, {Quirrenbach}, {Mitchell},
  {Albrecht}, {Hekker}, {Fischer}, {Marcy}, \& {Butler}}]{reffert06}
{Reffert}, S., {Quirrenbach}, A., {Mitchell}, D.~S., {et~al.} 2006, \apj, 652,
  661

\bibitem[{{Sato} {et~al.}(2003){Sato}, {Ando}, {Kambe}, {Takeda}, {Izumiura},
  {Masuda}, {Watanabe}, {Noguchi}, {Wada}, {Okada}, {Koyano}, {Maehara},
  {Norimoto}, {Okada}, {Shimizu}, {Uraguchi}, {Yanagisawa}, \&
  {Yoshida}}]{sato03}
{Sato}, B., {Ando}, H., {Kambe}, E., {et~al.} 2003, \apjl, 597, L157

\bibitem[{{Sato} {et~al.}(2012){Sato}, {Omiya}, {Harakawa}, {Izumiura},
  {Kambe}, {Takeda}, {Yoshida}, {Itoh}, {Ando}, {Kokubo}, \& {Ida}}]{sato12}
{Sato}, B., {Omiya}, M., {Harakawa}, H., {et~al.} 2012, \pasj, 64, 135

\bibitem[{{Setiawan} {et~al.}(2003){Setiawan}, {Hatzes}, {von der L{\"u}he},
  {Pasquini}, {Naef}, {da Silva}, {Udry}, {Queloz}, \& {Girardi}}]{seti03}
{Setiawan}, J., {Hatzes}, A.~P., {von der L{\"u}he}, O., {et~al.} 2003, \aap,
  398, L19

\bibitem[{{Skrutskie} {et~al.}(2006){Skrutskie}, {Cutri}, {Stiening},
  {Weinberg}, {Schneider}, {Carpenter}, {Beichman}, {Capps}, {Chester},
  {Elias}, {Huchra}, {Liebert}, {Lonsdale}, {Monet}, {Price}, {Seitzer},
  {Jarrett}, {Kirkpatrick}, {Gizis}, {Howard}, {Evans}, {Fowler}, {Fullmer},
  {Hurt}, {Light}, {Kopan}, {Marsh}, {McCallon}, {Tam}, {Van Dyk}, \&
  {Wheelock}}]{2mass}
{Skrutskie}, M.~F., {Cutri}, R.~M., {Stiening}, R., {et~al.} 2006, \aj, 131,
  1163

\bibitem[{{Tokovinin} \& {Smekhov}(2002)}]{toko02}
{Tokovinin}, A.~A. \& {Smekhov}, M.~G. 2002, \aap, 382, 118

\bibitem[{{van Leeuwen}(2007)}]{hip}
{van Leeuwen}, F. 2007, \aap, 474, 653

\bibitem[{{Villaver} \& {Livio}(2009)}]{vill09}
{Villaver}, E. \& {Livio}, M. 2009, \apjl, 705, L81

\bibitem[{{Vogt}(1987)}]{vogt87}
{Vogt}, S.~S. 1987, \pasp, 99, 1214

\bibitem[{{Vogt} {et~al.}(2002){Vogt}, {Butler}, {Marcy}, {Fischer},
  {Pourbaix}, {Apps}, \& {Laughlin}}]{vogt02}
{Vogt}, S.~S., {Butler}, R.~P., {Marcy}, G.~W., {et~al.} 2002, \apj, 568, 352

\bibitem[{{Walker} {et~al.}(1992){Walker}, {Bohlender}, {Walker}, {Irwin},
  {Yang}, \& {Larson}}]{walk92}
{Walker}, G.~A.~H., {Bohlender}, D.~A., {Walker}, A.~R., {et~al.} 1992, \apjl,
  396, L91

\bibitem[{{Wittenmyer} {et~al.}(2009){Wittenmyer}, {Endl}, {Cochran},
  {Ram{\'{\i}}rez}, {Reffert}, {MacQueen}, \& {Shetrone}}]{witt09}
{Wittenmyer}, R.~A., {Endl}, M., {Cochran}, W.~D., {et~al.} 2009, \aj, 137,
  3529

\bibitem[{{Wittenmyer} {et~al.}(2012){Wittenmyer}, {Horner}, {Tuomi}, {Salter},
  {Tinney}, {Butler}, {Jones}, {O'Toole}, {Bailey}, {Carter}, {Jenkins},
  {Zhang}, {Vogt}, \& {Rivera}}]{witt12}
{Wittenmyer}, R.~A., {Horner}, J., {Tuomi}, M., {et~al.} 2012, \apj, 753, 169

\bibitem[{{Zirm}(2007)}]{zirm07}
{Zirm}, H. 2007, IAU Commission on Double Stars, 161, 1

\end{thebibliography}

\listofobjects

\Online

\appendix

\onecolumn

\section{Maximum RV change for stars with companions}

In the case of a visual binary star system, one might not have any information
about the orbit other than the current separation.  In this case, one can
still calculate the maximum RV change of any orbit in that configuration, as
long as the distance to the system is known.

\begin{figure}
\centering
{\includegraphics{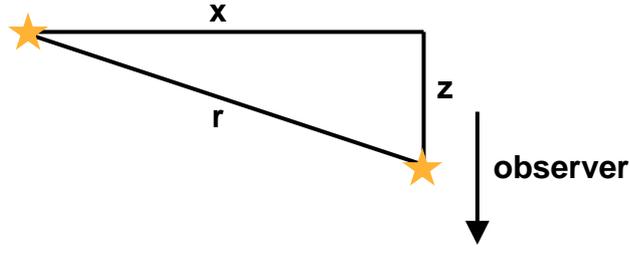}}
\caption{Diagram of binary star system.  The star of interest is on the right
  and the companion is on the left.  The observer is below the system in
  the plane of the page, so that the separation $r$ can be broken into the
  component along the line of sight, $z$, and the projection of the separation
  visible to the observer, $x$.}
\label{diagram}
\end{figure}

Figure~\ref{diagram} shows the orientation of our system.  The star and
companion are separated by distance $r$.  If we assume the observer is in
the plane of the page, then we can break that distance into the $x$ component,
the projection of the separation visible to the observer, and the $z$
component along the line of sight.  We know that the acceleration of
our star of interest, due to the gravitational interaction with its companion,
is

\begin{equation}
a=\frac{Gm_{c}}{r^{2}}
\end{equation}

where $G$ is the gravitational constant, $m_{c}$ is the companion mass, and
$r$ is the physical separation between them.  Since we do not know the
physical separation, but instead only have the projection of this separation,
we must find the acceleration in terms of the projected separation $x$ and the
component of the separation along our line of sight, $z$.  Here we are only
concerned with the acceleration along our line of sight, which is

\begin{equation}
a_{z}=\frac{Gm_{c}z}{(x^{2}+z^{2})^\frac{3}{2}}.
\end{equation}

To find the maximum value for the acceleration we must differentiate,

\begin{equation}
\frac{da_{z}}{dz}=\frac{Gm_{c}(x^{2}-2z^{2})}{(x^{2}+z^{2})^{\frac{5}{2}}},
\end{equation}

then set this equal to zero.  This gives

\begin{equation}
z_{max}=\frac{x}{\sqrt{2}}
\end{equation}

which is the condition at the maximum acceleration.  Inserting this back into
Equation~A.2, and putting the projected separation $x$ in terms of the
observed angular separation $\rho$ and the parallax $\varpi$, gives the maximum
acceleration along the line of site

\begin{equation}
a_{z_{max}}=\frac{2Gm_{c}\varpi^{2}}{3\sqrt{3}\rho^{2}}
\end{equation}

for any orbit of a star with a companion of given mass.


\section{RV data}


\begin{longtable}{lrr}
\caption{RV data for \taugem}
\label{tgdata}\\
\hline\hline\noalign{\smallskip}
JD\tablefootmark{a} [days] & $v_{\mathrm{rad}}$ [\ms] & $\sigma_{\mathrm{RV}}$ [\ms]\\
\hline\noalign{\smallskip}
\endhead
\hline\noalign{\smallskip}
\endfoot

 1809.039 & $ -277.8 $ &  3.6 \\
 1854.888 & $  -12.6 $ &  4.0 \\
 1898.902 & $  223.8 $ &  6.5 \\
 1930.810 & $  334.5 $ &  4.4 \\
 1990.659 & $  167.6 $ &  5.6 \\
 2012.653 & $   -4.7 $ &  4.0 \\
 2013.651 & $  -37.5 $ &  4.7 \\
 2014.656 & $  -20.8 $ &  4.0 \\
 2175.047 & $   94.1 $ &  3.7 \\
 2176.055 & $  136.4 $ &  4.3 \\
 2177.035 & $  124.9 $ &  3.6 \\
 2192.995 & $  200.7 $ &  4.5 \\
 2193.967 & $  220.5 $ &  3.8 \\
 2205.938 & $  280.7 $ &  4.5 \\
 2206.949 & $  270.7 $ &  4.6 \\
 2222.898 & $  329.3 $ &  4.5 \\
 2258.811 & $  332.9 $ &  5.1 \\
 2259.853 & $  299.8 $ &  5.1 \\
 2295.773 & $  115.8 $ &  5.8 \\
 2297.842 & $  131.0 $ &  6.3 \\
 2307.730 & $   98.9 $ &  5.0 \\
 2336.741 & $ -129.0 $ &  4.8 \\
 2361.708 & $ -300.2 $ &  4.4 \\
 2362.721 & $ -271.6 $ &  4.5 \\
 2363.688 & $ -293.6 $ &  4.8 \\
 2384.667 & $ -327.7 $ &  4.3 \\
 2393.654 & $ -354.9 $ &  6.1 \\
 2529.039 & $  336.2 $ &  4.4 \\
 2531.009 & $  317.8 $ &  4.6 \\
 2532.039 & $  326.1 $ &  4.3 \\
 2542.018 & $  330.5 $ &  4.7 \\
 2543.981 & $  339.0 $ &  4.7 \\
 2560.013 & $  334.8 $ &  6.8 \\
 2562.004 & $  322.0 $ &  4.6 \\
 2571.969 & $  284.3 $ &  4.6 \\
 2574.010 & $  265.6 $ &  4.9 \\
 2589.975 & $  201.6 $ &  4.8 \\
 2590.892 & $  205.7 $ &  4.5 \\
 2603.925 & $  149.2 $ &  5.3 \\
 2604.916 & $  115.8 $ &  5.1 \\
 2605.867 & $   99.1 $ &  5.1 \\
 2615.842 & $   47.0 $ &  4.5 \\
 2616.858 & $   48.9 $ &  4.5 \\
 2627.816 & $  -63.7 $ &  3.8 \\
 2665.755 & $ -290.7 $ &  5.2 \\
 2667.815 & $ -328.3 $ &  4.9 \\
 2699.687 & $ -375.4 $ &  5.1 \\
 2717.691 & $ -314.8 $ &  4.1 \\
 2765.676 & $  -49.3 $ & 18.3 \\
 2900.034 & $  178.6 $ &  4.6 \\
 2932.987 & $  -87.4 $ &  4.2 \\
 2934.975 & $ -105.0 $ &  4.3 \\
 2964.028 & $ -314.4 $ &  5.2 \\
 2966.896 & $ -321.8 $ &  4.3 \\
 2985.944 & $ -350.6 $ &  5.3 \\
 3022.849 & $ -358.7 $ &  5.6 \\
 3089.704 & $   81.9 $ &  4.8 \\
 3111.665 & $  229.1 $ &  4.1 \\
 3271.015 & $ -263.2 $ &  3.6 \\
 3357.929 & $ -164.1 $ &  4.2 \\
 3400.811 & $  131.6 $ &  4.6 \\
 3424.711 & $  254.7 $ &  4.3 \\
 3444.674 & $  322.0 $ &  5.5 \\
 3651.003 & $ -227.0 $ &  4.1 \\
 3701.974 & $  103.2 $ &  5.1 \\
 3741.888 & $  268.5 $ &  4.1 \\
 3790.746 & $  310.0 $ &  4.7 \\
 3857.654 & $ -124.8 $ &  4.8 \\
 4054.935 & $  344.2 $ &  4.4 \\
 4057.940 & $  315.3 $ &  4.6 \\
 4123.835 & $  126.0 $ &  4.5 \\
 4182.662 & $ -250.5 $ &  5.3 \\
 4227.649 & $ -373.4 $ &  4.6 \\
 4392.004 & $  353.7 $ &  6.6 \\
 4419.958 & $  231.7 $ &  4.9 \\
 4480.917 & $ -210.0 $ &  4.6 \\
 4504.767 & $ -317.3 $ &  4.9 \\
 4754.975 & $   41.5 $ &  3.8 \\
 4807.924 & $ -266.5 $ &  4.6 \\
 4882.830 & $ -217.6 $ &  6.3 \\
 4911.674 & $   -8.9 $ &  5.7 \\
 5102.055 & $ -243.4 $ &  6.4 \\
 5122.025 & $ -315.3 $ &  5.5 \\
 5154.966 & $ -334.6 $ &  7.3 \\
 5242.708 & $  142.0 $ &  6.1 \\
 5282.681 & $  348.6 $ &  5.4 \\
 5464.063 & $ -267.5 $ &  6.9 \\
 5569.854 & $  245.8 $ &  4.8 \\
 5593.771 & $  341.2 $ &  4.9 \\
 5620.740 & $  331.8 $ &  5.7 \\
 5651.688 & $  179.7 $ &  5.1 \\
 5830.017 & $   28.2 $ &  8.9 \\
 5863.951 & $  253.6 $ &  4.0 \\
 5865.031 & $  267.5 $ &  4.1 \\
 5894.944 & $  341.1 $ &  6.1 \\

\hline\noalign{\smallskip}
\tablefoottext{a}{Julian date--2450000}
\end{longtable}


\begin{longtable}{lrr}
\caption{RV data for 91~Aqr}
\label{91aqdata}\\
\hline\hline\noalign{\smallskip}
JD\tablefootmark{a} [days] & $v_{\mathrm{rad}}$ [\ms] & $\sigma_{\mathrm{RV}}$ [\ms]\\
\hline\noalign{\smallskip}
\endhead
\hline\noalign{\smallskip}
\endfoot

 1744.976 & $   72.1 $ &  3.9 \\
 1780.886 & $   60.9 $ &  4.3 \\
 1807.843 & $  -46.3 $ &  4.8 \\
 1853.694 & $  -75.3 $ &  4.7 \\
 1856.737 & $  -76.5 $ &  4.6 \\
 1899.645 & $   65.6 $ &  5.0 \\
 1932.598 & $  101.3 $ &  5.5 \\
 2049.000 & $  -54.6 $ &  5.7 \\
 2079.999 & $   50.9 $ &  5.1 \\
 2083.991 & $   82.4 $ &  4.2 \\
 2098.992 & $   65.9 $ &  5.0 \\
 2099.962 & $   71.3 $ &  4.7 \\
 2100.982 & $   71.2 $ &  4.7 \\
 2107.959 & $   86.8 $ &  5.2 \\
 2109.955 & $   60.0 $ &  4.2 \\
 2125.939 & $   82.6 $ &  4.7 \\
 2163.837 & $  -10.0 $ &  4.9 \\
 2164.896 & $    5.7 $ &  5.1 \\
 2165.868 & $    4.5 $ &  4.6 \\
 2166.885 & $   -0.7 $ &  5.4 \\
 2174.843 & $  -45.9 $ &  5.2 \\
 2175.820 & $  -56.4 $ &  5.8 \\
 2176.798 & $  -55.5 $ &  4.7 \\
 2177.826 & $  -42.3 $ &  5.7 \\
 2192.772 & $  -75.8 $ &  5.0 \\
 2193.805 & $  -95.2 $ &  5.0 \\
 2194.786 & $ -110.7 $ &  4.4 \\
 2205.714 & $  -92.8 $ &  5.6 \\
 2206.754 & $ -103.4 $ &  5.1 \\
 2207.716 & $  -98.8 $ &  5.3 \\
 2222.754 & $  -89.1 $ &  5.7 \\
 2223.698 & $  -81.4 $ &  5.5 \\
 2258.635 & $   31.2 $ &  6.2 \\
 2259.675 & $   44.9 $ &  5.8 \\
 2295.594 & $  111.2 $ &  6.3 \\
 2297.595 & $  116.8 $ &  7.0 \\
 2437.985 & $   45.5 $ &  6.0 \\
 2438.963 & $   59.6 $ &  5.7 \\
 2439.990 & $   30.8 $ &  5.2 \\
 2452.995 & $   82.0 $ &  5.5 \\
 2453.944 & $   55.5 $ &  5.6 \\
 2454.958 & $   49.3 $ &  6.2 \\
 2464.987 & $   72.1 $ &  5.3 \\
 2465.954 & $   44.9 $ &  5.4 \\
 2466.975 & $   80.3 $ &  5.3 \\
 2471.924 & $   53.6 $ &  5.6 \\
 2483.979 & $  102.5 $ &  5.2 \\
 2484.947 & $   88.1 $ &  5.4 \\
 2494.978 & $   94.5 $ &  6.0 \\
 2495.941 & $   94.7 $ &  5.8 \\
 2496.970 & $  101.0 $ &  5.2 \\
 2505.951 & $   65.0 $ &  6.3 \\
 2506.919 & $   62.9 $ &  5.4 \\
 2507.851 & $   31.3 $ &  6.1 \\
 2517.914 & $   13.5 $ &  6.3 \\
 2519.944 & $   -4.1 $ &  5.8 \\
 2528.862 & $  -18.8 $ &  5.5 \\
 2529.917 & $  -27.8 $ &  6.7 \\
 2530.887 & $  -44.4 $ &  5.3 \\
 2531.897 & $  -47.3 $ &  5.2 \\
 2532.858 & $  -34.6 $ &  5.7 \\
 2541.884 & $  -72.1 $ &  5.5 \\
 2542.825 & $  -66.1 $ &  6.2 \\
 2543.849 & $  -53.8 $ &  5.4 \\
 2559.876 & $  -91.9 $ &  6.0 \\
 2560.792 & $  -96.1 $ &  6.0 \\
 2571.766 & $  -85.2 $ &  5.9 \\
 2573.779 & $  -64.9 $ &  6.0 \\
 2589.758 & $  -75.6 $ &  6.0 \\
 2590.812 & $  -55.8 $ &  6.1 \\
 2603.749 & $   -8.9 $ &  6.1 \\
 2604.645 & $  -16.5 $ &  5.7 \\
 2605.685 & $  -34.7 $ &  6.1 \\
 2615.702 & $   43.7 $ &  5.7 \\
 2616.731 & $   48.8 $ &  6.4 \\
 2617.620 & $   18.7 $ &  6.6 \\
 2801.984 & $   -2.7 $ &  5.7 \\
 2803.980 & $   15.7 $ &  5.4 \\
 2837.959 & $  113.0 $ &  6.9 \\
 2838.979 & $  114.0 $ &  7.1 \\
 2839.976 & $   71.4 $ &  5.9 \\
 2861.929 & $   94.8 $ &  4.4 \\
 2863.972 & $   98.8 $ &  5.3 \\
 2879.906 & $   42.9 $ &  5.6 \\
 2881.945 & $    0.5 $ &  9.5 \\
 2898.891 & $  -19.1 $ &  5.2 \\
 2899.847 & $  -50.3 $ &  5.6 \\
 2900.895 & $  -62.0 $ &  5.2 \\
 2932.829 & $  -75.3 $ &  5.5 \\
 2934.777 & $ -103.2 $ &  4.7 \\
 2963.743 & $  -81.8 $ &  5.4 \\
 2966.731 & $  -43.3 $ &  6.0 \\
 3022.615 & $  103.6 $ &  6.4 \\
 3232.884 & $   24.7 $ &  4.1 \\
 3234.951 & $   16.6 $ &  4.7 \\
 3265.858 & $  -74.9 $ &  3.9 \\
 3268.852 & $  -79.3 $ &  4.5 \\
 3286.776 & $  -90.8 $ &  3.8 \\
 3355.642 & $   55.1 $ &  5.0 \\
 3546.976 & $   61.2 $ &  4.5 \\
 3578.894 & $   99.0 $ &  5.6 \\
 3578.929 & $   80.0 $ &  5.4 \\
 3578.997 & $   55.2 $ &  5.4 \\
 3579.849 & $   81.1 $ &  4.6 \\
 3579.946 & $   62.1 $ &  4.3 \\
 3579.995 & $   63.7 $ &  4.1 \\
 3580.021 & $   65.0 $ &  4.1 \\
 3580.859 & $   70.2 $ &  4.9 \\
 3580.939 & $   70.4 $ &  4.5 \\
 3581.021 & $   73.1 $ &  4.8 \\
 3581.825 & $   74.7 $ &  4.5 \\
 3581.926 & $   69.7 $ &  4.2 \\
 3582.022 & $   59.2 $ &  4.4 \\
 3582.849 & $   55.1 $ &  4.6 \\
 3582.915 & $   54.7 $ &  4.7 \\
 3583.008 & $   69.1 $ &  4.8 \\
 3583.931 & $   80.6 $ &  4.2 \\
 3583.939 & $   74.3 $ &  4.0 \\
 3583.948 & $   74.9 $ &  4.1 \\
 3584.822 & $   67.7 $ &  4.4 \\
 3584.870 & $   66.0 $ &  4.2 \\
 3584.943 & $   59.3 $ &  4.1 \\
 3585.821 & $   38.4 $ &  3.9 \\
 3585.924 & $   82.6 $ &  4.0 \\
 3586.010 & $   89.9 $ &  4.3 \\
 3586.825 & $   70.3 $ &  4.1 \\
 3586.852 & $   64.9 $ &  4.3 \\
 3587.018 & $   60.7 $ &  4.0 \\
 3587.921 & $   15.6 $ &  4.2 \\
 3587.985 & $   35.9 $ &  5.3 \\
 3698.663 & $  -27.6 $ &  4.7 \\
 3702.650 & $    3.4 $ &  5.5 \\
 3741.621 & $   86.1 $ &  5.5 \\
 3911.975 & $  107.0 $ &  4.3 \\
 3915.984 & $   77.1 $ &  4.1 \\
 3934.930 & $   98.3 $ &  4.3 \\
 3936.910 & $   95.1 $ &  4.6 \\
 3967.907 & $  -20.2 $ &  4.5 \\
 3982.797 & $  -15.9 $ &  4.4 \\
 4054.746 & $  -18.9 $ &  5.1 \\
 4124.594 & $   66.1 $ &  4.4 \\
 4266.991 & $   74.7 $ &  4.0 \\
 4297.882 & $  134.7 $ &  4.7 \\
 4300.885 & $  107.3 $ &  5.2 \\
 4314.872 & $   92.0 $ &  3.9 \\
 4344.849 & $  -19.6 $ &  4.6 \\
 4348.875 & $  -32.3 $ &  4.7 \\
 4391.656 & $  -35.8 $ &  6.9 \\
 4419.688 & $   -3.5 $ &  4.8 \\
 4421.725 & $   -4.4 $ &  5.5 \\
 4482.612 & $   78.9 $ &  6.4 \\
 4645.956 & $   88.1 $ &  5.3 \\
 4667.963 & $   75.9 $ &  5.0 \\
 4711.856 & $  -60.2 $ &  5.1 \\
 4756.777 & $  -95.1 $ &  4.8 \\
 4777.699 & $  -22.7 $ &  5.1 \\
 5026.951 & $  110.8 $ &  7.3 \\
 5063.950 & $    7.2 $ &  5.7 \\
 5097.906 & $  -90.0 $ &  6.8 \\
 5116.818 & $  -36.6 $ &  6.3 \\
 5155.655 & $   60.3 $ &  8.1 \\
 5363.976 & $   74.8 $ &  5.4 \\
 5420.911 & $  -19.9 $ &  4.9 \\
 5449.762 & $  -63.3 $ &  5.0 \\
 5569.617 & $   73.5 $ &  5.1 \\
 5704.007 & $   -4.1 $ &  4.7 \\
 5732.971 & $  104.9 $ &  4.6 \\
 5735.972 & $   77.3 $ &  4.8 \\
 5757.920 & $   43.1 $ &  5.7 \\
 5803.845 & $  -25.8 $ &  5.3 \\
 5806.857 & $  -67.3 $ &  5.3 \\
 5829.792 & $  -90.3 $ &  6.0 \\
 5862.718 & $  -66.0 $ &  4.8 \\
 5894.680 & $   27.1 $ &  7.9 \\

\hline\noalign{\smallskip}
\tablefoottext{a}{Julian date--2450000}
\end{longtable}


\begin{longtable}{lrr}
\caption{CRIRES data for \taugem}
\label{tgcrires}\\
\hline\hline\noalign{\smallskip}
JD\tablefootmark{a} [days] & $v_{\mathrm{rad}}$ [\ms] & $\sigma_{\mathrm{RV}}$ [\ms]\\
\hline\noalign{\smallskip}
\endhead
\hline\noalign{\smallskip}
\endfoot

 5859.712 & $  203.0 $ & 40.2 \\
 5918.848 & $  323.6 $ & 40.2 \\
 5994.624 & $    2.4 $ & 40.2 \\
 6039.424 & $ -292.8 $ & 40.2 \\
 6244.736 & $  252.1 $ & 40.2 \\
 6288.768 & $ -101.9 $ & 40.2 \\
 6326.400 & $ -316.5 $ & 40.2 \\

\hline\noalign{\smallskip}
\tablefoottext{a}{Julian date--2450000}
\end{longtable}


\begin{longtable}{lrr}
\caption{CRIRES data for 91~Aqr}
\label{91acrires}\\
\hline\hline\noalign{\smallskip}
JD\tablefootmark{a} [days] & $v_{\mathrm{rad}}$ [\ms] & $\sigma_{\mathrm{RV}}$ [\ms]\\
\hline\noalign{\smallskip}
\endhead
\hline\noalign{\smallskip}
\endfoot

 5854.592 & $ -116.3 $ & 40.1 \\
 6112.896 & $   98.4 $ & 40.1 \\
 6144.640 & $   69.1 $ & 40.1 \\
 6239.616 & $  -11.1 $ & 40.1 \\

\hline\noalign{\smallskip}
\tablefoottext{a}{Julian date--2450000}
\end{longtable}


\end{document}